\documentclass[twocolumn]{aastex62}
\usepackage{bm}
\usepackage{amsmath}
\renewcommand{\vec}[1]{\bm{#1}}

\begin{document}

\title{The Fluid-like and Kinetic Behavior of Kinetic Alfv\'{e}n Turbulence in Space Plasma}
 

\author[0000-0002-0786-7307]{Honghong Wu}
\affiliation{School of Earth and Space Sciences, Peking University, Beijing, China}
\affiliation{Mullard Space Science Laboratory, University College London, Surrey, UK}

\author{Daniel Verscharen}
\affiliation{Mullard Space Science Laboratory, University College London, Surrey, UK}
\affiliation{Space Science Center, University of New Hampshire, Durham NH, USA} 

\author{Robert T. Wicks}
\affiliation{Mullard Space Science Laboratory, University College London, Surrey, UK}
\affiliation{Institute for Risk and Disaster Reduction, University College London, London, UK}

\author{Christopher H. K. Chen} 
\affiliation{School of Physics and Astronomy, Queen Mary University of London, London, UK} 

\author{Jiansen He}
\affiliation{School of Earth and Space Sciences, Peking University, Beijing, China} 

\author{Georgios Nicolaou}
\affiliation{Mullard Space Science Laboratory, University College London, Surrey, UK} 

\begin{abstract} 

Kinetic Alfv\'{e}n waves (KAWs) are the short-wavelength extension of the MHD Alfv\'{e}n-wave branch in the case of highly-oblique propagation with respect to the background magnetic field. Observations of space plasma show that small-scale turbulence is mainly KAW-like. We apply two theoretical approaches, collisional two-fluid theory and collisionless linear kinetic theory, to obtain predictions for the KAW polarizations depending on $\beta_\mathrm{p}$ (the ratio of the proton thermal pressure to the magnetic pressure) at the ion gyroscale in terms of fluctuations in density, bulk velocity, and pressure. We perform a wavelet analysis of MMS magnetosheath measurements and compare the observations with both theories. We find that the two-fluid theory predicts the observations better than kinetic theory, suggesting that the small-scale KAW-like fluctuations exhibit a fluid-like behavior in the magnetosheath although the plasma is weakly collisional. We also present predictions for the KAW polarizations in the inner heliosphere that are testable with Parker Solar Probe and Solar Orbiter.
 
\end{abstract}

\keywords{plasmas, waves, turbulence, magnetohydrodynamics (MHD), solar-terrestrial relations}


\section{Introduction} \label{sec:intro} 
Standard single-fluid magnetohydrodynamics (MHD) contains four linear modes in a collisional plasma: the Alfv\'{e}n wave, the fast-magnetosonic wave, the slow-magnetosonic wave, and the entropy mode. The polarization and dispersion relations identify their counterparts in collisionless kinetic theory \citep{Stix1992book, Gary1993book}. Due to the collisionless nature of the solar wind, kinetic models have been expected to describe these fluctuations more accurately than MHD. However, \cite{Verscharen2017ApJ} compare theoretical predictions for wave polarization properties from both MHD theory and kinetic theory with observations of compressive fluctuations in the solar wind and find that the predictions from linear MHD agree better with the observations than the predictions from kinetic theory for slow modes at large scales. The comparison between the fluid-like and the kinetic behavior of collisionless plasmas is of great importance for our fundamental understanding of plasma turbulence. In this context, we define a fluid-like mode as a plasma mode that follows the predictions from fluid equations with adiabatic and isotropic pressure closure.   

The magnetic fluctuations in the inertial range of solar wind turbulence exhibit Alfv\'{e}nic correlations \citep{Belcher1971JGR} and a scale-dependent anisotropy with $k_{\perp}\gg k_{\parallel}$ \citep{Horbury2008PhRvL, Wicks2010MNRAS, Chen2011MNRAS, Chen2012ApJ, He2013ApJ, Yan2016ApJL}, where $k_{\perp}$ is the perpendicular wavenumber and $k_{\parallel}$ is the parallel wavenumber with respect to the background magnetic field. Kinetic Alfv\'{e}n waves (KAWs) are the short-wavelength extension of the Alfv\'{e}n-wave branch in the case of highly-oblique propagation with respect to the background magnetic field. Therefore, it is thought that the cascade continues into the KAW-like regime for $k_{\perp}\rho_\mathrm{p} \gtrsim 1$, where $\rho_\mathrm{p}=v_\mathrm{th}/\omega_{\mathrm{cp}}$ is the ion gyroscale, $v_\mathrm{th} $ is the perpendicular thermal speed, $\omega_{c\mathrm{p}}=q_\mathrm{p}B_0/m_\mathrm{p}c$ is the proton gyrofrequency, $q_\mathrm{p}$ is the proton electric charge, $m_\mathrm{p}$ is the proton mass, $B_0$ is the magnitude of the background magnetic field, and $c$ is the speed of light. A growing body of evidence corroborates the presence of kinetic Alfv\'{e}n turbulence in the solar wind \citep{Chandran2009ApJ, Sahraoui2010PhRvL, Chen2010PhRvL, He2011ApJ, Salem2012ApJ, Chen2013PhRvL}. Likewise, observations of small-scale fluctuations in the magnetosheath also suggest the presence of KAW-like turbulence \citep{Chen2017PhRvL, Breuillard2018ApJ}.  
We extend Verscharen et al.'s (\citeyear{Verscharen2017ApJ}) study to examine KAW-like fluctuations at small scales. In Section \ref{sec:Theory}, we present predictions for fluctuations in the first three velocity moments associated with KAWs from both collisional two-fluid theory and collisionless linear kinetic theory. In Section \ref{sec:Data Analysis}, we describe our analysis of MMS magnetosheath measurements. We compare observations with our predictions in Section \ref{sec:Results}. In Section \ref{Discussion and Conclusions}, we discuss our results and present our conclusions. We present predictions for solar-wind measurements with Parker Solar Probe and Solar Orbiter in the Appendix. 
 
\section{Theory} \label{sec:Theory}
We take the background magnetic field to be $\vec{B}_0=(0,0,B_0)$, the electric field to be $ \vec{E}=\delta\vec{ E}$ ($\delta\vec{ E}$ is the fluctuating electric field) and the wave vector to be $\vec{k}=(k_{\perp},0,k_{\parallel})$. In our coordinate system, the $z$-direction is parallel to $\vec B_0$, and $\vec k$ lies in the $x-z$ plane. We define the dimensionless quantities $\xi_s$, $\chi_{s\parallel }$, $\chi_{s\perp }$ and $\psi_s$ as the normalized amplitudes of the fluctuations in the first three velocity moments of species $s$ through 
\begin{equation} 
\frac{\delta n_s}{n_{0s}}=\xi_s \frac{\delta B_\mathrm{y}}{B_0}, 
\end{equation}
\begin{equation}  
\frac{\delta v_{\parallel s}}{v_\mathrm{A}}=\chi_{\parallel s}\frac{\delta B_\mathrm{y}}{B_0}, 
\end{equation} 
\begin{equation} 
\frac{\delta v_{\mathrm{y}s}}{v_\mathrm{A}}=\chi_{\perp s} \frac{\delta B_\mathrm{y}}{B_0},
\end{equation}  and 
\begin{equation}
\frac{\delta P_s}{P_{B0}}=\psi_s \frac{\delta B_\mathrm{y}}{B_0}, 
\end{equation}  
where $P_{B0}=B_0^2/8\pi$; $n_{0s}$ is the average density of species $s$, $\delta n_s$, $ \delta v_{\parallel s}$, $ \delta v_\mathrm{ys}$, $\delta P_s$, and $ \delta B_\mathrm{y}$ are the amplitudes of fluctuations in the number density, the bulk velocity component parallel to $\vec{B}_0$, the bulk velocity component in the $y$-direction, the thermal pressure, and the magnetic field component in the $y$-direction, respectively. Note that $\xi_s$, $\chi_{\parallel s}$, $\chi_{\perp s}$ and $\psi_s$ are complex and include imformation about the phases between the fluctuating quantities. We consider a plasma consisting of protons ($s=\mathrm{p}$) and electrons ($s=\mathrm{e}$) only and neglect all effects of temperature anisotropies. 

\subsection{Two-fluid Model} 
The two-fluid dispersion relation follows from linearizing the continuity, momentum, and energy equations and Maxwell's equations. We rewrite Hollweg's (\citeyear{Hollweg1999JGR}) two-fluid solutions for the dispersion relation of KAWs using our coordinate system. The KAW dispersion relation is then given by
\begin{multline}
\omega^2 =\frac{k_{\parallel}^2v_\mathrm{A}^2}{\displaystyle  2\left(1+G+ k_{\perp}^2d_\mathrm{p}^2\frac{m_\mathrm{e}}{m_\mathrm{p}}\right)} \left[1+2G+Gk_{\perp}^2d_\mathrm{p}^2+\right.\\
\left. \left[(1+Gk_{\perp}^2d_\mathrm{p}^2)^2+4Gk_{\perp}^2d_\mathrm{p}^2\left(G-\frac{m_\mathrm{e}}{m_\mathrm{p}}\right)\right]^{1/2}\right],
\end{multline}  
where $v_\mathrm{A}=B_0/\sqrt{4\pi n_\mathrm{p} m_\mathrm{p}}$ is the Alfv\'{e}n speed, $d_\mathrm{p}=v_\mathrm{A}/\omega_{\mathrm{cp}}$ is the ion inertial length, $G=(\gamma_\mathrm{p}\beta_\mathrm{p}+\gamma_\mathrm{e}\beta_\mathrm{e})/2$, 
\begin{equation}
\beta_s=\frac{8\pi n_s\kappa_\mathrm{B} T_s}{ B_0^2 },
\end{equation} 
$\gamma_s$ is the specific heat ratio of species $s$, $m_\mathrm{e}$ is the electron mass, $n_s$ is the number density of species $s$, $T_s$ is the parallel temperature of species $s$, and $\kappa_\mathrm{B}$ is Boltzmann constant.
 
We calculate the quantities $\xi_\mathrm{p}$, $\chi_{\parallel \mathrm{p}}$, $\chi_{\perp \mathrm{p}}$, and $\psi_\mathrm{p}$ as  
\begin{multline} 
\xi_\mathrm{p} =\frac{\displaystyle 2i ( \omega^2  -k_{\parallel}^2v_\mathrm{A}^2)}{ \displaystyle k_{\perp}d_\mathrm{p}[\omega^2 (2m_\mathrm{e}/m_\mathrm{p}- \gamma_\mathrm{p}\beta_\mathrm{p} )-  \gamma_\mathrm{e}\beta_\mathrm{e} k_{\parallel}^2v_\mathrm{A}^2]}\\
\times \frac{ \displaystyle \omega}{\displaystyle v_\mathrm{A}\left(k_{\parallel}-k_{\perp}\frac{\delta E_\mathrm{z}}{\delta E_\mathrm{x}}\right)},
\end{multline}  
\begin{multline}
\chi_{\parallel \mathrm{p}} =\left[\frac{i \omega_\mathrm{cp} }{\omega}+\frac{i \gamma_\mathrm{p}\beta_\mathrm{p} \omega_\mathrm{cp}k_{\parallel}^2v_\mathrm{A}^2 }{\omega ( \gamma_\mathrm{e}\beta_\mathrm{e}k_{\parallel}^2v_\mathrm{A}^2 -2 \omega^2m_\mathrm{e}/m_\mathrm{p} )} \right]\frac{\delta E_\mathrm{z}}{\delta E_\mathrm{x}} \\ 
\times \frac{ \displaystyle  \omega}{\displaystyle v_\mathrm{A}\left(k_{\parallel}-k_{\perp}\frac{\delta E_\mathrm{z}}{\delta E_\mathrm{x}} \right)},
\end{multline}    
\begin{multline}
\chi_{\perp \mathrm{p}} = \left[-1-\frac{ \displaystyle i \omega \frac{\delta E_\mathrm{y}}{\delta E_\mathrm{x}}}{ \displaystyle  \omega_\mathrm{cp} } -\frac{ \displaystyle  \gamma_\mathrm{p}\beta_\mathrm{p}  k_{\perp}k_{\parallel} v_\mathrm{A}^2\frac{\delta E_\mathrm{z}}{\delta E_\mathrm{x}}}{ \gamma_\mathrm{e}\beta_\mathrm{e} k_{\parallel}^2v_\mathrm{A}^2-2\omega^2 m_\mathrm{e}/m_\mathrm{p} }\right]\\ 
\times \frac{ \displaystyle \omega}{\displaystyle v_\mathrm{A}\left(k_{\parallel}-k_{\perp}\frac{\delta E_\mathrm{z}}{\delta E_\mathrm{x}}\right)},
\end{multline}  
and 
\begin{equation}
\psi_\mathrm{p} ={\gamma_\mathrm{p}\beta_\mathrm{p}}\xi_\mathrm{p},
\end{equation} 
where    
\begin{multline}
 \frac{\delta E_\mathrm{y}}{\delta E_\mathrm{x}}=\frac{-i\omega \omega_\mathrm{cp}(\gamma_\mathrm{p}\beta_\mathrm{p}+\gamma_\mathrm{e}\beta_\mathrm{e})}{  \left[ \omega^2 ( \gamma_\mathrm{p}\beta_\mathrm{p} -2m_\mathrm{e}/m_\mathrm{p})+ \gamma_\mathrm{e}\beta_\mathrm{e} k_{\parallel}^2v_\mathrm{A}^2\right] }\\
\times \frac{ \omega^2-k_{\parallel}^2v_\mathrm{A}^2 }{ \omega^2-k_{\parallel}^2v_\mathrm{A}^2-k_{\perp}^2v_\mathrm{A}^2 }
\end{multline} 
and
\begin{equation}
 \frac{\delta E_\mathrm{z}}{\delta E_\mathrm{x}}=\frac{(\omega^2-k_{\parallel}^2v_\mathrm{A}^2)( \gamma_\mathrm{e}\beta_\mathrm{e} k_{\parallel}^2v_\mathrm{A}^2-2 \omega^2m_\mathrm{e}/m_\mathrm{p})}{k_{\parallel}k_{\perp}v_\mathrm{A}^2\left[ \omega^2(2m_\mathrm{e}/m_\mathrm{p}- \gamma_\mathrm{p}\beta_\mathrm{p})- \gamma_\mathrm{e}\beta_\mathrm{e} k_{\parallel}^2v_\mathrm{A}^2\right]}.
\end{equation}  
     
\subsection{Kinetic Theory} 
The linear kinetic hot-plasma dispersion relation follows from linearizing the Vlasov equation and Maxwell's equations \citep[see, e.g.,][Chapter 10]{Stix1992book}. We assume that both protons and electrons have isotropic Maxwellian background distribution functions. The dispersion relation is then given by the non-trival solutions to the wave equation,
\begin{equation}
\frac{\vec{k}c}{\omega} \times \left(\frac{\vec{k}c}{\omega} \times \delta \vec{E} \right)+\epsilon \delta \vec{E} \equiv \vec{D}\delta \vec{E} = 0,
\end{equation}\label{equation12}
where $\epsilon$ is the dielectric tensor. We note that $\omega$ is complex. The electric current density fulfills
\begin{equation}
\vec{j}= \sum_s \vec{j}_s=-\frac{i\omega}{4\pi}\sum \vec{\sigma}_s \cdot \delta \vec{E},
\end{equation} 
where $\vec{j}_s$ is the current density contribution from species $s$, $\vec{\sigma}_s$ is the susceptibility of species $s$, and $\epsilon=1+\sum_s \vec{\sigma}_s$.

The linearized continuity equation connects the fluctuations of the density with the fluctuations of the current density,
\begin{equation}
-\omega q_s n_{0s} \delta n_s+\vec{k} \cdot \vec{j}_s=0.
\end{equation}
Therefore, we find 
\begin{equation} 
\xi_s = \frac{-iB_0}{4\pi n_{0s}^2q_s}\vec{k} \cdot \left( \vec{\sigma}_s \cdot \frac{\delta \vec{E}}{\delta B_\mathrm{y}}\right).
\end{equation} 
The electric current density also fulfills
\begin{equation}
\vec{j}= \sum_s \vec{j}_s=\sum_s q_s\int \vec{v}_s\, \delta f_s \,d^3\vec{v},
\end{equation} 
leading to
\begin{equation}
\chi_{\parallel s}  =\frac{-i\omega B_0 }{4\pi n_{0s} q_s v_\mathrm{A} }\left( \vec{\sigma}_s \cdot \frac{\delta \vec{E}}{\delta B_\mathrm{y}}\right)_\mathrm{z}
\end{equation}  
and
\begin{equation}
\chi_{\perp s}  =\frac{-i\omega B_0 }{4\pi n_{0s} q_s v_\mathrm{A} }\left( \vec{\sigma}_s \cdot \frac{\delta \vec{E}}{\delta B_\mathrm{y}}\right)_\mathrm{y}.
\end{equation}   
The fluctuations in pressure follow from the second moment of the fluctuations in the distribution $\delta f_s$,
\begin{equation}
\psi_{\parallel s} = \frac{n_sm_\mathrm{p}B_0}{2P_{B0} \,\delta B_\mathrm{y}} \int  v_{\parallel s}^2\delta f_s \,d^3v
\end{equation}
and 
\begin{equation}
\psi_{\perp s} =\frac{n_sm_\mathrm{p}B_0}{2P_{B0} \,\delta B_\mathrm{y}} \int  v_{\perp s}^2\delta f_s \,d^3v.
\end{equation} 
We define fluctuations in the total pressure as
\begin{equation}
\psi_s = \frac{1}{3}\psi_{\parallel s}+ \frac{2}{3}\psi_{\perp s}.
\end{equation}
 
\begin{figure}[ht!]
\includegraphics[width=\linewidth]{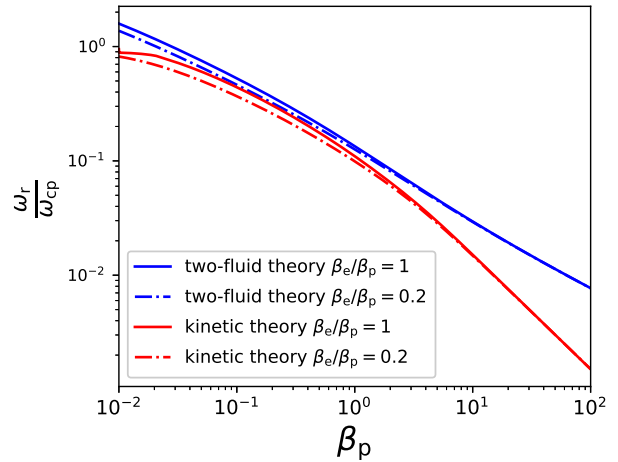}
\caption{Real part of the frequency of KAWs with $k_{\perp}\rho_\mathrm{p}=2$ as a function of $\beta_\mathrm{p}$ from the two-fluid theory (blue) and kinetic theory (red). We use $\theta_\mathrm{kB}=88^{\circ}$, $\gamma_\mathrm{p}=5/3$ and $\gamma_\mathrm{e}=1$, where $\theta_\mathrm{kB}$ is the angle between the direction of wave propagation and the background magnetic field. The solid and dash-dot lines represent $\beta_\mathrm{e}/\beta_\mathrm{p}=1$ and $\beta_\mathrm{e}/\beta_\mathrm{p}=0.2$ respectively. Two-fluid theory and kinetic theory agree quantitively. The value of $\beta_\mathrm{e}/\beta_\mathrm{p}$ affects the predictions.} \label{fig:figure1}
\end{figure}

The NHDS (New Hampshire Dispersion relation Solver) code \citep{Verscharen2018RNA} solves the hot-plasma dispersion relation of the linearized Vlasov-Maxwell system and determines polarization properties numerically. Figure \ref{fig:figure1} shows NHDS solutions for $\omega_\mathrm{r}$ (the real part of the frequency) of KAWs with $k_{\perp}\rho_\mathrm{p}=2$ as functions of $\beta_\mathrm{p}$ for $\beta_\mathrm{e}/\beta_\mathrm{p}=1$ and $\beta_\mathrm{e}/\beta_\mathrm{p}=0.2$. In addition, we show KAW solutions from our two-fluid theory with the same parameters and $\gamma_\mathrm{p}=5/3$ and $\gamma_\mathrm{e}=1$. The dispersion relations agree quantitively well, especially for $\beta_\mathrm{p}\lesssim 1$.
  
\section{Data Analysis} \label{sec:Data Analysis}
We analyze data from the four MMS spacecraft \citep{Burch2016SSR}. In burst mode, the Fast Plasma Investigation (FPI) instrument provides proton moments every 150 ms and electron moments every 30 ms \citep{Pollock2016SSR}. The Fluxgate Magnetometer (FGM) measures the magnetic field with a resolution $<1$ ms. We select intervals in the magnetosheath from 01 October 2015 to 28 February 2018 based on MMS's quicklook archive. We split every interval into 10-second intervals with an overlap of 5 seconds. We remove data intervals in which $\max[|n_\mathrm{p}-n_\mathrm{e}|/n_\mathrm{p}]>0.1$.
 
Magnetosheath plasma exhibits fluctuations not related to pure turbulence, e.g., instability-generated waves (such as mirror modes, ion-cyclotron waves, and whistler waves) and various other non-turbulent structures \citep[see, e.g.,][]{Lucek2005SSRv}. In order to investigate the turbulence itself, we eliminate intervals with $\max(\beta_\mathrm{p})-\min(\beta_\mathrm{p})> \beta_{0\mathrm{p}} $ and $\max(n_\mathrm{p})-\min(n_\mathrm{p})>0.2\,n_{0\mathrm{p}}$, where the subscript $0$ represents the average over the 10-second interval. We also remove data intervals in which $v_{0\mathrm{p}}<v_\mathrm{A}$, where $v_{0\mathrm{p}}$ is the average ion bulk velocity so that Taylor's hypothesis applies \citep{Taylor1938RS, Klein2014ApJL}. For each interval, we calculate $\beta_\mathrm{0p}$ and $\beta_\mathrm{0e}$. We find that the average ratio $\beta_\mathrm{0e}/\beta_\mathrm{0p}\approx 0.21$ with a standard deviation of 0.09 and, therefore, set $\beta_\mathrm{e}/\beta_\mathrm{p}=0.2$ in our study. The averaged normalized amplitude of the fluctuations in $\delta B_{\perp}$ is $\delta B_{\perp}/B_0=0.05$ with a standard deviation of $0.04$, and the averaged normalized amplitude of the fluctuations in magnetic-field amplitude is $\delta |\vec B|/B_0=0.07$ with a standard deviation of $0.05$. The Columb collision frequency $\nu_\mathrm{ei}$ is $7 \times10^{-5} $ $\mathrm{Hz}$ under the conditions in our dataset: $ n_\mathrm{e} \approx 25 $ $\mathrm{ cm}^{-3}$, $T_\mathrm{e}\approx 50 $ $\mathrm{eV}$, corresponding to a mean free path $\lambda\approx2 \times10^7 $ $\mathrm{km}$ for $V_\mathrm{0p} \approx 150 $ $\mathrm{km/s}$. Since $\lambda/d_\mathrm{p}\approx 1500$, we consider the plasma to be collisionless. 
 
In order to study the scale-dependent behavior of the fluctuations, we apply a continuous wavelet transform based on the Morlet wavelet \citep{Torrence1998Compo} to $n_\mathrm{p}$, all components of $\vec{v}_\mathrm{p}$ and $\vec{B}$, and to every element of the proton pressure tensor $\vec{P}_\mathrm{p}$. We obtain the flucuation amplitudes $\delta n_\mathrm{p}$, $\delta \vec{V}_\mathrm{p}$, $\delta \vec{B}$, $\delta \vec{P}_\mathrm{p}$ from the absolute wavelet coefficients as functions of both scale $\ell$ ($16$ logarithmically spaced scales) and time $t$.  We calculate the local magnetic field and the local velocity at wavelet scale $\ell$ and time $t_n$ by weighting the time series with a Gaussian curve centered at time $t_n$ \citep{Horbury2008PhRvL, Podesta2009ApJ} as  
\begin{equation}
\vec{C}_n = \frac{\displaystyle \sum_{m=0}^{N-1} \vec{C}_m \exp\left[-\frac{(t_n-t_m)^2}{2\ell^2}\right]}{\displaystyle \sum_{m=0}^{N-1} \exp\left[-\frac{(t_n-t_m)^2}{2\ell^2}\right]},
\end{equation} 
where $N$ is the number of data points and $ \vec{C}$ is either $ \vec{B}$ or $ \vec{V}$. In this way, we obtain the local magnetic field $\vec{B}_0$, the local velocity $\vec{V}_{0\mathrm{p}}$, the angle between velocity and local magnetic field $\theta_{VB}$, as well as $\beta_\mathrm{p}$ and $\rho_\mathrm{p}$ as functions of $\ell$ and $t_n$. 

We find that $\delta B_\mathrm{y}\gg\delta B_\mathrm{x}$ for KAWs in both two-fluid theory and kinetic theory. Therefore, we use $\delta B_\mathrm{\perp}\approx\delta B_\mathrm{y}$ and exploit the azimuthal symmetry of a gyrotropic distribution of the fluctuating energy. We calculate the parallel and perpendicular fluctuations by  
\begin{equation}
\delta \vec{C}_{\parallel}=\frac{|\delta \vec{C} \cdot \vec{B}_0 |}{| \vec{B}_0 |} 
\end{equation}    
and
\begin{equation}
  \delta \vec{C}_{\perp}=\frac{|\vec{B}_0 \times(\delta \vec{C} \times \vec{B}_0) |}{| \vec{B}_0 |^2},
\end{equation}  
where $\delta \vec{C}$ is either $\delta \vec{V}_\mathrm{p}$ or $\delta \vec{B}$.

We obtain the transformation matrix $\alpha_{ij}$ from the Geocentric Solar Ecliptic (GSE) coordinate system to the field-aligned coordinate system with respect to $\vec{B}_0$ and then transform the tensor of pressure fluctuations $\delta P_{\mathrm{p}ij}$ to the field-aligned coordinate system by
\begin{equation}
\delta P'_{\mathrm{p}ij}= \alpha_{ki} \, \delta P_{\mathrm{p}km} \, \alpha_{mj}.
\end{equation}  
The total proton thermal pressure is given by 
\begin{equation} 
\delta P_\mathrm{p} =\frac{1}{3}(\delta P'_{\mathrm{p}11}+ \delta P'_{\mathrm{p}22} + \delta P'_{\mathrm{p}33} ) .
\end{equation} 
  
For each scale $\ell$ and time $t_n$, we determine $k_{\perp}$ as
\begin{equation} 
k_{\perp}\rho_\mathrm{p}=\frac{ 2\pi \rho_\mathrm{p}} {\ell\,v_{0\mathrm{p}} \sin(\theta_{VB})},
\end{equation} 
and retain only coefficients with $1.8\leq k_{\perp}\rho_\mathrm{p}\leq 2.2$ and $60^o<\theta_{VB}<120^o$.
 
In order to remove noise in our measurements, we only retain coefficients when $\delta v_{\parallel\mathrm{p} }>1.0\, \mathrm{km/s} $ and $\delta v_{ \perp\mathrm{p}}>1.0 \, \mathrm{km/s} $. We then construct a set of 600 bins in logarithmic space of $\beta_\mathrm{p}$ and the value of the coefficients, count the number of coefficients located at each bin, and column-normalize the number of counts by the bin length of the coefficients value.
    
\section{Results} \label{sec:Results}
\begin{figure}[ht!]
\includegraphics[width=\linewidth]{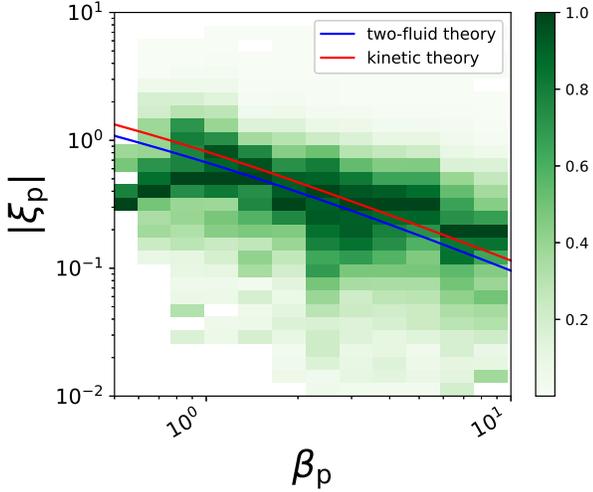}
\caption{Ratio of proton density fluctuations and perpendicular magnetic field fluctuations, $\xi_\mathrm{p}$, as a function of $\beta_\mathrm{p}$ at $k_{\perp}\rho_i\approx 2$. The lines show our theoretical results and the color-coded bins show the logarithmically scaled, column-normalized data distribution in the $\xi_\mathrm{p}-\beta_\mathrm{p}$ plane. The theoretical predictions from two-fluid theory and kinetic theory are similar, and the observations agree with both models.}\label{fig:figure2}
\end{figure}

Figure \ref{fig:figure2} shows the zeroth proton velocity moment $\xi_\mathrm{p}$ depending on $\beta_\mathrm{p}$ in two-fluid and kinetic theory and our MMS observations for $ k_{\perp}\rho_\mathrm{p}=2$. Two-fluid theory and kinetic theory predict similar values for $\xi_\mathrm{p}$ under these parameters. Therefore, the observations of the zeroth moment do not favor the applicability of either theory unambiguously. However, the data agree well with the theoretical predictions as well as previous observations \citep{Chen2013PhRvL, Chen2017PhRvL} and thus support the applicability of our data analysis technique and the model comparisons.

\begin{figure}[ht!]
\includegraphics[width=\linewidth]{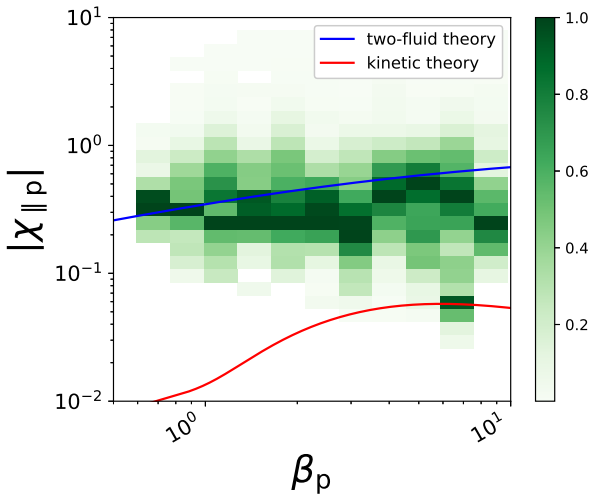}
\caption{Ratio of parallel proton velocity fluctuations and perpendicular magnetic field fluctuations, $\chi_{\parallel \mathrm{p}}$, as a function of $\beta_\mathrm{p}$ at $k_{\perp}\rho_\mathrm{p}\approx 2$ with the same panel and line styles as Figure \ref{fig:figure2}. The theoretical predictions for $\chi_{\parallel \mathrm{p} }$ from two-fluid theory are significantly greater than those from kinetic theory, and the observations agree better with two-fluid theory than with kinetic theory.}\label{fig:figure3}
\end{figure}

Figure \ref{fig:figure3} shows the first parallel proton velocity moment $\chi_{\parallel \mathrm{p}}$ depending on $\beta_\mathrm{p}$ in both theories and our MMS observations. For all shown values of $\beta_{\mathrm p}$, $\chi_{\parallel \mathrm{p}}$ is significantly greater in our results from two-fluid theory than in our results from kinetic theory. Two-fluid theory predicts a value that is approximately the observed value of $\chi_{\parallel \mathrm p}$. Kinetic theory, on the other hand underestimates the observed value by a factor between about three and thirty in the shown range of $\beta_{\mathrm p}$. This observation suggests that the plasma exhibits fluid-like behavior at $k_{\perp}\rho_\mathrm{p} \approx 2$.  
 
\begin{figure}[ht!]
\includegraphics[width=\linewidth]{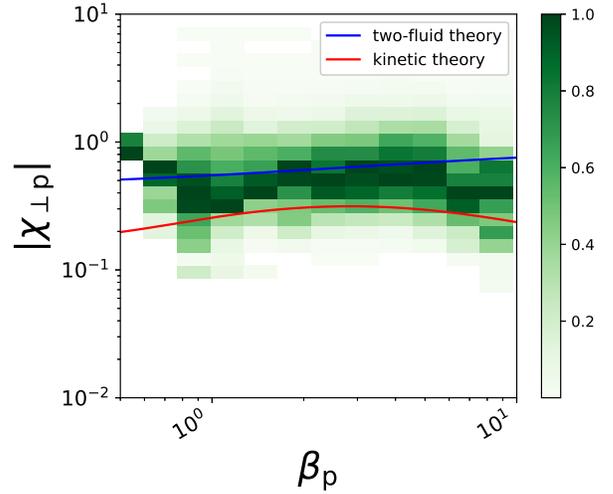}
\caption{Ratio of perpendicular proton velocity fluctuations and perpendicular magnetic field fluctuations, $\chi_{\perp \mathrm{p}}$, as a function of $\beta_\mathrm{p}$ at $k_{\perp}\rho_\mathrm{p}\approx 2$ with the same panel and line styles as Figure \ref{fig:figure2}. The theoretical predictions for $\chi_{\perp \mathrm{p}}$ from two-fluid theory are greater than the predictions from kinetic theory, and the observations agree with both two-fluid theory and kinetic theory with a small bias toward two-fluid theory.}\label{fig:figure4}
\end{figure}

Figure \ref{fig:figure4} shows the first perpendicular proton velocity moment $\chi_{\perp \mathrm{p}}$ depending on $\beta_\mathrm{p}$ in both theories and our MMS observations. The predictions from our two-fluid model are greater than the predictions from kinetic theory by a factor of about three. The observations lie between both predictions with a small bias toward two-fluid theory.

\begin{figure}[ht!]
\includegraphics[width=\linewidth]{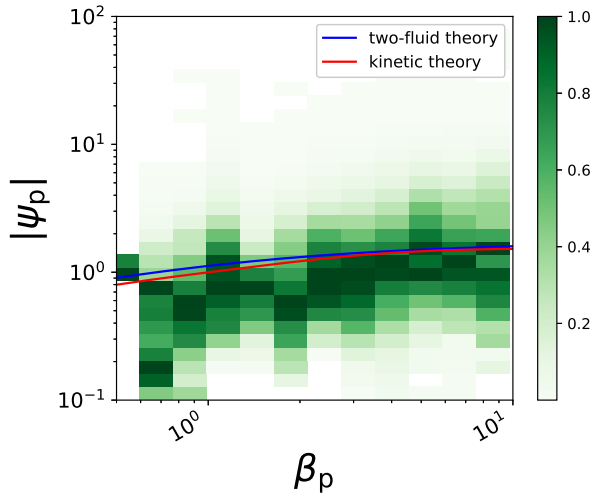}
\caption{Ratio of ion pressure fluctuations and perpendicular magnetic field fluctuations, $\psi_\mathrm{p}$, as a function of $\beta_\mathrm{p}$ at $k_{\perp}\rho_\mathrm{p}\approx 2$ with the same panel and line styles as Figure \ref{fig:figure2}. The theoretical predictions for $\psi_\mathrm{p}$ from two-fluid theory and kinetic theory are similar, and the observations agree with both models.}\label{fig:figure5}
\end{figure}

Figure \ref{fig:figure5} shows the second proton velocity moment $\psi_\mathrm{p}$ depending on $\beta_\mathrm{p}$ in both theories and our MMS observations. Like in Figure \ref{fig:figure2} for $\xi_{\mathrm p}$, two-fluid theory and kinetic theory predict similar values for $\psi_\mathrm{p}$, and both predict the observations well. 

We note that an adjustment of $\gamma_\mathrm{p}$ and $\gamma_\mathrm{e}$ within reasonable values does not lead to a significant change in our results. Moreover, an adjustment of $\theta_{kB}$ within the range of $85^{\circ}\lesssim \theta_{kB}<90^{\circ}$ does not lead to a significant change in our results, confirming our assumption that the observed fluctuations are indeed consistent with highly-oblique KAW turbulence. 

\section{Discussion and Conclusions} \label{Discussion and Conclusions}
The behavior of KAWs departs from the behavior of Alfv\'{e}n waves mainly due to two effects. The ion motion is affected by compression and introduces a polarization-drift term in the equation of motion. Furthermore, the parallel component of the electric field is non-zero and electrons move in the field-parallel direction to neutralize the ion density perturbations. We derive predictions for the proton polarization of KAWs using collisional two-fluid theory and collisionless kinetic theory. In the linear kinetic theory, the fluctuations are represented by fluctuations in the distribution function, $\delta f_s$ so that all moments are included and generally non-zero. Our two-fluid theory, on the other hand, assumes an adiabatic and isotropic closure for the moment hierarchy, i.e., there are no fluctuations in heat flux and the pressure tensor is isotropic. Apart from these differences, fluid and kinetic theory are equivalent. Measurements with even higher velocity-space resolution may be capable of showing the heat-flux suppression in the future.

Both theories predict similar behaviors for density and pressure fluctuations, but the parallel and perpendicular velocity fluctuations show clear differences: these fluctuations are greater in two-fluid theory than in kinetic theory. Due to the noise in the velocity observation, we cannot rule out the possibility that fluctuations with very small amplitude exhibit a behavior consistent with kinetic theory. However, our comparison of fluctuations in the magnetosheath above the noise level with our theoretical predictions shows that KAW turbulence behaves fluid-like at ion scales, suggesting that some of the fluid-like behavior found by \cite{Verscharen2017ApJ} extends to the ion-scale fluctuations. We note that relaxing our assumption of temperature isotropy may improve the agreement between our theory and observations. In addition, a study based on a superposition of KAW turbulence with other modes at small scales may modify our results since our present method does not distinguish the contributions from different wave types than KAWs. A comparison with alternative approximations to the dispersion relation \citep{Hunana2013ApJ, Sulem2015JPlPh, Told2016NJPh} may give further insight into the physics of the observed modes. However, these extensions are beyond the scope of this work. 
 
Our finding of fluid-like behavior in KAW turbulence suggests that some yet unknown mechanism creates conditions similar to the adiabatic and isotropic closure applied in our two-fluid theory, even at small scales and under collisionless conditions. Anti-phase-mixing \citep{Schekochihin2016JPP} is a potential explanation for this fluid-like behavior. In the turbulent background, nonlinear interactions between fluctuations at different scales can trigger stochastic plasma echoes \citep{Gould1967PRL, Schekochihin2016JPP} that may inhibit the transfer of power to higher moments of the velocity distribution. \cite{Parker2017PoP} and \cite{Meyrand2018arxiv} found that energy transfer from large to small velocity-space scales nearly cancels due to “anti-phase-mixing” excited by a stochastic plasma echo. This process leads to an effective low-moment closure, even under collisionless conditions. In KAWs with larger amplitude, the nonlinear trapping of electrons may contribute to the saturation of damping and a more fluid-like behavior \citep{Gershman2017NatureC}.

Alternatively, wave-particle interactions can suppress fluctuations in higher moments of the velocity distribution. \cite{Verscharen2016ApJ} find that microinstabilities generate fluctuations that scatter protons and thus reduce the anisotropy of the pressure tensor. Wave-particle interactions may then play the role of particle-particle collisions in suppressing fluctuations in higher moments and closing the moment hierarchy at low order. 

Our finding of the fluid-like behavior of KAW turbulence at scales down to the proton inertial length supports the use of fluid models when studying large- and small-scale fluctuations. This discovery will be beneficial to astrophysical modeling since fluid computations are much faster than kinetic computations. More fundamentally, it is of great importance to determine the physics processes that lead to this fluid-like behavior of an otherwise collisionless plasma.

\acknowledgments 
Honghong Wu is supported by the China Scholarship Council for her stay at MSSL. This work is also supported by NSFC under contracts 41474147, 41674171 and 41574168, STFC Ernest Rutherford Fellowships ST/P003826/1 and ST/N003748/2, STFC Consolidated Grant ST/N000722/1 and STFC Solar Orbiter UK Community Support Grant ST/P005489/1. We appreciate valuable conversations with Lloyd Woodham, Owen Roberts, Alexander Pit\v{n}a, Petr Hellinger, Olga Alexandrova and Francesco Valentini. Honghong Wu is grateful for Chuanyi Tu's support during this work.

\appendix
\section{Predictions for kinetic Alfv\'{e}n turbulence in the inner heliosphere} 
In the near future, the space missions Parker Solar Probe and Solar Orbiter will carry instruments into the inner heliosphere that will provide us with unprecedented measurements of the fluctuations in both the fields and the particle distributions. Their data will allow us to study kinetic Alfv\'{e}n turbulence in the solar wind using our methods for the first time.

\begin{figure}[ht!]
\includegraphics[width=\linewidth]{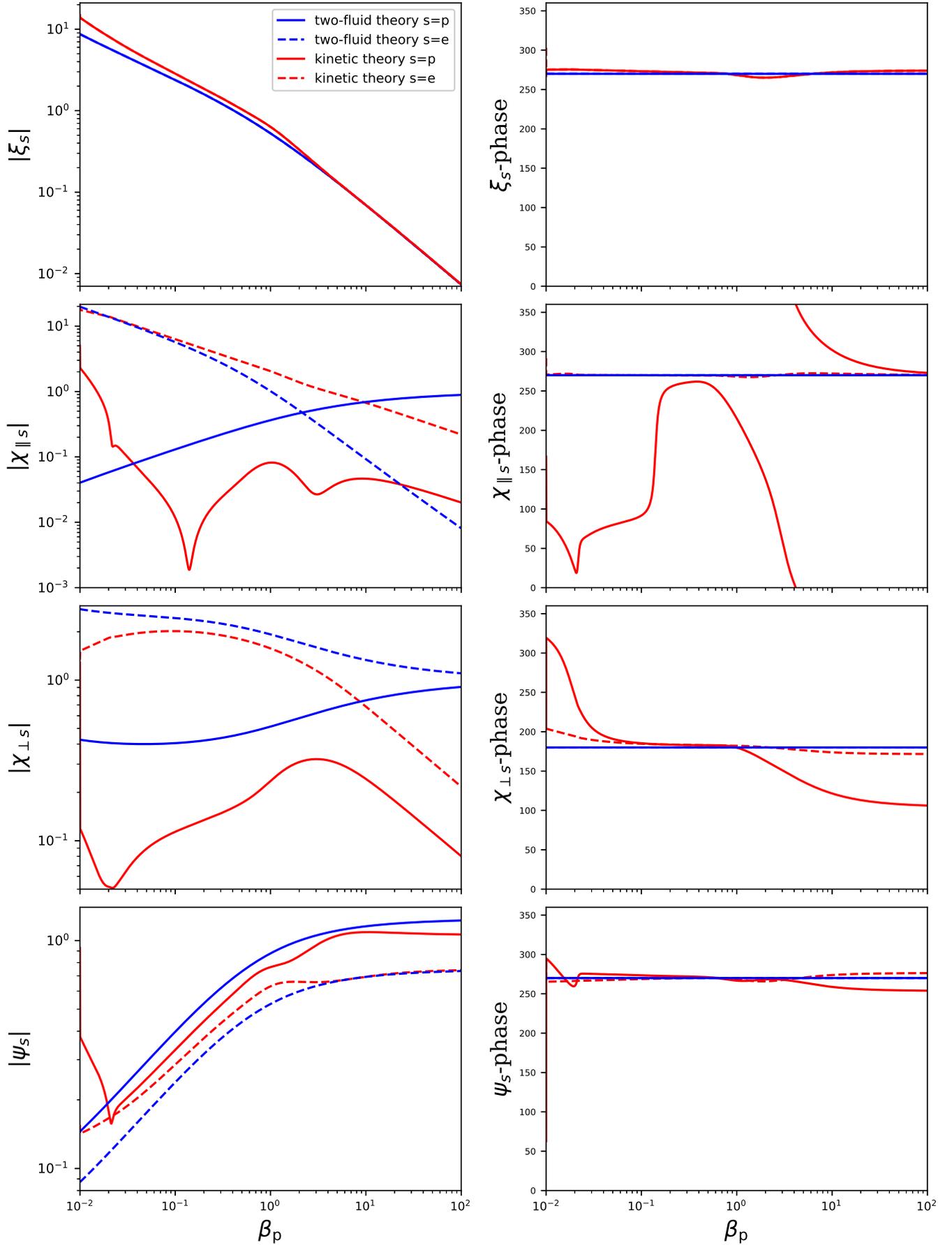}
\caption{Predictions of KAWs as a function of $\beta_\mathrm{p}$ at $k_{\perp}\rho_\mathrm{p}=2$. From top to bottom, these panels show $\xi_s$, $\chi_{\parallel s}$, $\chi_{\perp s}$ and $\psi_s$. The blue (red) lines represent two-fluid theory (kinetic theory). The left panels show the amplitudes and the right panels show the phases. The solid (dashed) lines represent protons (electrons).}\label{fig:figure6}
\end{figure}

The solar wind in the inner heliosphere exhibits a broader range of $\beta_\mathrm{p}$-values and typically shows $\beta_\mathrm{p}\approx \beta_\mathrm{e}$. We derive predictions for the KAW polarizations under typical solar-wind conditions with $\theta=88^{\circ}$, $T_{\parallel s}=T_{\perp s}$, $\beta_\mathrm{e}/\beta_\mathrm{p}=1.0$, $\gamma_\mathrm{p}=5/3$ and $\gamma_\mathrm{e}=1$. For completeness, we add the polarizations for electrons. Our kinetic theory applies to protons and electrons likewise. In two-fluid theory, the calculations for the electron polarizations differ from the calculations for the proton polarization. We find for electrons
\begin{equation}
\xi_\mathrm{e} =\xi_\mathrm{p},  
\end{equation}

\begin{equation}
\chi_{\parallel \mathrm{e}}=\left[-\frac{i \omega_\mathrm{cp} }{\omega}+\frac{i \gamma_\mathrm{e}\beta_\mathrm{e} \omega_\mathrm{cp}k_{\parallel}^2v_\mathrm{A}^2 }{\omega ( \gamma_\mathrm{e}\beta_\mathrm{e}k_{\parallel}^2v_\mathrm{A}^2 -2 \omega^2m_\mathrm{e} /m_\mathrm{p} )} \right] \frac{ \displaystyle  (m_\mathrm{p}/m_\mathrm{e})\,\,\,\omega\, \frac{\delta E_\mathrm{z}}{\delta E_\mathrm{x}}}{\displaystyle v_\mathrm{A}\left(k_{\parallel}-k_{\perp}\frac{\delta E_\mathrm{z}}{\delta E_\mathrm{x}} \right)},
\end{equation}

\begin{equation}
\chi_{\perp \mathrm{e}}=\left[-1 +\frac{ \displaystyle  \gamma_\mathrm{e}\beta_\mathrm{e}  k_{\perp}k_{\parallel} v_\mathrm{A}^2\frac{\delta E_\mathrm{z}}{\delta E_\mathrm{x}}}{  \gamma_\mathrm{e}\beta_\mathrm{e} k_{\parallel}^2v_\mathrm{A}^2-2\omega^2m_\mathrm{e}/m_\mathrm{p}  }\right]  \frac{ \displaystyle \omega}{\displaystyle v_\mathrm{A}\left(k_{\parallel}-k_{\perp}\frac{\delta E_\mathrm{z}}{\delta E_\mathrm{x}}\right)},
\end{equation} 
and
\begin{equation}
\psi_\mathrm{e} ={\gamma_\mathrm{e}\beta_\mathrm{e}}\xi_\mathrm{e}.
\end{equation}  
Figure \ref{fig:figure6} shows $\xi_s$, $\chi_{\parallel s}$, $\chi_{\perp s}$, and $\psi_s$ at $k_{\perp}\rho_\mathrm{p}=2$ as defined in Section \ref{sec:Theory}. Under solar-wind conditions, two-fluid theory and kinetic theory predict similar $\xi_s$ and $\psi_s$ (in both amplitude and phase) behaviors depending on $\beta_\mathrm{p}$. For $\chi_{\parallel s}$ and $\chi_{\perp s}$, our theoretical results show large differences in both amplitude and phase.

We expect Parker Solar Probe and Solar Orbiter to test our predictions based on large statistical data sets in the solar wind at different distances from the Sun. This future study will improve our understanding of fluctuations at ion scales and the differences between kinetic- and fluid-like behavior in the solar wind.
 


\end{document}